\documentclass[useAMS,usenatbib,usegraphicx]{mn2e}
\usepackage{times}

\newif\ifAMStwofonts
\AMStwofontstrue

\newcommand{\simlt}{\lower.5ex\hbox{$\; \buildrel < \over \sim \;$}}
\newcommand{\simgt}{\lower.5ex\hbox{$\; \buildrel > \over \sim \;$}}
\newcommand{\be}{\begin{equation}}
\newcommand{\ba}{\begin{eqnarray}}
\newcommand{\ee}{\end{equation}}
\newcommand{\ea}{\end{eqnarray}}

\title[Role of Environment on Early-type Galaxies]
{The Role of Environment on the Formation of Early-Type Galaxies}
\author[B. Rogers et al.]
{Ben Rogers$^1$, Ignacio Ferreras$^2$\thanks{E-mail: ferreras@star.ucl.ac.uk},
Anna Pasquali$^3$, Mariangela Bernardi$^4$, Ofer Lahav$^5$,
\newauthor
and Sugata Kaviraj$^{2,6,7}$\\
$^1$ Department of Physics, King's College London, Strand, London WC2R 6LS\\
$^2$ Mullard Space Science Laboratory, University College London, 
Holmbury St Mary, Dorking, Surrey RH5 6NT\\
$^3$ Max-Planck-Institut f\"ur Astronomie, K\"onigstuhl 17, D-69117 Heidelberg, Germany\\
$^4$ Department of Physics and Astronomy, University of Pennsylvania, USA\\
$^5$ Department of Physics and Astronomy, University College London,
  Gower St. London WC1E 6BT\\
$^6$ Blackett Laboratory, Imperial College London, London SW 2AZ\\
$^7$ Astronomy group, The Denys Wilkinson Building, Keble Road, Oxford, OX1 3RH\\
}

\begin{document}
\date{MNRAS in press -- arXiv version}
\pagerange{\pageref{firstpage}--\pageref{lastpage}} \pubyear{2010}
\maketitle
\label{firstpage}

\begin{abstract}
The effect of environment on galaxy formation poses one of the best
constraints on the interplay between mass assembly and star formation
in galaxies. We present here a detailed study of the stellar
populations of a volume-limited sample of early-type galaxies from the
{\sl Sloan Digital Sky Survey}, across a range of environments --
defined as the mass of the host dark matter halo, according to the
groups catalogue of Yang et~al. The stellar populations are explored
through the SDSS spectra, via projection onto a set of two spectral
vectors determined from Principal Component Analysis. This method has
been found to highlight differences not seen when using standard,
model-dependent comparisons of photo-spectroscopic data. We find the
velocity dispersion of the galaxy to be the main driver behind
the different star formation histories of early-type
galaxies. However, environmental effects are seen to play a role
(although minor).  Our Principal Components allow us to
  distinguish between the effects of environment as a change in
  average age (mapping the time lapse of assembly) or the presence of
  recent star formation (reflecting environment-related
  interactions).  Galaxies populating the lowest mass halos have
stellar populations on average $\sim$1~Gyr younger than the rest of
the sample. The fraction of galaxies with small amounts of recent star
formation is also seen to be truncated when occupying halos more
massive than M$_{\rm H}\simgt 3\times 10^{13}$M$_\odot$. The sample is
split into satellite and central galaxies for a further analysis of
environment. Small but measurable differences are found between these
two subsamples. For an unbiased comparison, we have to restrict this
analysis to a range of halo masses over which a significant number of
central and satellite galaxies can be found. Over this mass range,
satellites are {\sl younger} than central galaxies of the same stellar
mass. The younger satellite galaxies in M$_{\rm H}\sim 6\times
10^{12}$M$_\odot$ halos have stellar populations consistent with the
central galaxies found in the lowest mass halos of our sample
(i.e. M$_{\rm H}\sim 10^{12}$M$_\odot$).  This result is indicative of
galaxies in lower mass halos being accreted into larger halos.
\end{abstract}


\section{Introduction}

The origin and evolution of early-type galaxies is a long debated
topic, its solution involving a large array of cosmological and
astrophysical processes. The current paradigm of galaxy formation is
embedded in the $\Lambda$CDM cosmology, from which the structures in
the Universe are built hierarchically. Under this framework,
early-type galaxies are effectively a secondary stage in galaxy
evolution. The first stage consists of the formation of
rotationally-supported disk galaxies, built up through the accretion
of gas and smaller systems. Mergers subsequently operate, creating
early-type galaxies. The masses and ages of the stellar populations in
early-type galaxies imply that these systems are the result of an
intense starburst, followed by processes which quench star formation
after which the galaxy evolves passively.

One of the proposed mechanisms to stop star formation requires the
removal of gas from the galaxy, involving either fast removal of cold
gas (ram-pressure stripping), or a slower removal of hot diffuse gas
(strangulation). These mechanisms, however, do not result in a change
in kinematics and only minor changes in morphology \citep{wein09} and
so they can possibly explain the increased fraction of S0 galaxies
\citep[e.g.][]{drez80}. The major formation process of red sequence
galaxies is thought to operate through major mergers
\cite[e.g.][]{delucia06} which result in the required structural and
dynamical changes. Such processes have been known for a considerable
time to produce spheroidal galaxies \citep{toomre72,BarnHern96,KB03},as
well as more detailed photometric and dynamical properties
\citep{naab03,naab06}. More recent results also suggest that
minor mergers may play an increasingly important role in the build up
and size evolution of massive ellipticals at relatively later times
\citep[e.g.][]{ks06,bezan09, bern09}. The subsequent quenching of
star formation and evolution of the galaxy onto the red sequence,
requires the gas to be removed or heated to prevent the formation of
new stars. Currently models invoke feedback from active
galactic nuclei (AGN), since this fits naturally into the merger
scenario. Such interactions are expected to drive material to the
centre of the galaxy through tidal torques, towards the central
supermassive black hole (SMBH) \citep{dimat07}. The discovery of a
correlation between the mass of the SMBH and galaxy mass \citep{geb00},
put significant weight behind the idea and provided scope for more
comprehensive theories \citep{hopkins06,faber07,somerville08}.

This scenario naturally introduces an expectation of environmental
dependence, since such formation process involves interactions with
neighbouring galaxies and structures. A correlation with environment
can arise in two forms. Firstly through the initial conditions, as
these provide the impetus for the formation of the first galaxies so
that objects in dense environments will form earlier than in average
or low density regions \citep{gottlober01,berlind03}. Secondly, in
higher density regions interactions, mergers, gas stripping, etc, are
more likely to take place over the lifetime of a galaxy and so
galaxies in these environments will be pushed onto the red sequence at
earlier times. Certainly the fact that red early-type galaxies are
preferentially found in higher density environments
\citep[e.g. ][]{drez80,blanton05,wein06}, suggests that environment
plays an important role in their evolution. Therefore, looking at
environmental differences in the stellar populations of early-type
galaxies offers a method by which to constrain their formation.

There has been much work on studies of differences in stellar
populations through many different methods; the variations in the
tight correlations followed by the early-type population such as the
colour magnitude relation \cite[see e.g.][]{gallazzi06} and the
fundamental plane \cite[see e.g.][]{bern03}, galaxy colours \cite[see
e.g.][]{blanton05}, absorption line indices \cite[see e.g.][]{nelan05}
and the parameters of population synthesis modelling \cite[see
e.g.][]{bern06,thomas05}. In all cases the effect of environment has
been shown to be relatively weak, if observed at all. Thus we take a
different approach, choosing a different methodology involving
principal component analysis on spectral data to identify small
differences between the stellar populations of early-type galaxies
\citep{pca}, in a similar style to \cite{igpca}, over a range of
environments.

The environment is in most cases quantified through the projected
number density of galaxies, typically the distance to the n$^{th}$
nearest neighbour. However it has been argued that more physically
motivated scales of environment are the mass and the virial radius of
the host dark matter halo \citep{kauff04,yng05,wein06,blanton07} .
Not only are environmental dependencies observed to act primarily over
distances comparable to the virial radius of such halos
\citep{goto03}, but also the merger history of the dark matter halo is
determined mainly by its present mass \citep{kauff04}. The mass of the
host dark matter halo cannot be directly measured in most cases but
can be estimated through galaxy group catalogues
\cite[e.g.][]{yng07}. Such catalogues also provide the halo-centric
radius and can be used to easily separate the sample into central and
satellite populations. The other advantage in estimating the
environment through halo mass is that it allows a direct comparison
between observations and theoretical models. \citet{dekel06},
\citet{somerville08}, \citet{catt08} or \citet{KO_08} all make model 
predictions of the evolution and properties of galaxies in terms of 
the dark matter halo mass. For example \cite{catt08}, following the 
work of \citet{birn03} and \citet{dekel06}, suggested that the downsizing
observed in elliptical galaxies can be modelled by considering a
critical mass halo above which gas cannot be accreted efficiently,
being shock heated to the virial temperature, thus effectively
shutting down star formation.

This paper is structured as follows: we describe the sample of
early-type galaxies used in this study as well as the details of the
principal component analysis. We investigate the results of the PCA
projections over a range of halo mass and velocity dispersion, we highlight the
differences observed and investigate them using the stellar population
models of \cite{BC03}. The sample is finally split into central and
satellite galaxies, whose properties are compared. The satellite
population is then used to determine a possible dependence on
halo-centric radius.

\section{The sample}

This work is based on the large sample of early-type galaxies
of \citet{pca}. This sample is selected from the \citet{bern06}
catalogue, compiled from the Sloan Digital Sky Survey
\citep[SDSS, ][]{SDSStec}, Data Release 4 \citep{dr4}. It is a
volume-limited sample within z$\leq$0.1 and $M_r\leq -$21. A cut with
respect to  signal
to noise ratio (S/N) was also imposed, rejecting those spectra with
S/N$\leq$15. The final sample comprises 7,134 early type
galaxies. Here we extend \cite{pca} to investigate the effect of
environment in more detail, including the information of the host dark
matter halo for each galaxy, a explained below.

\begin{figure}
  \begin{center}
    \includegraphics[width=3.3in]{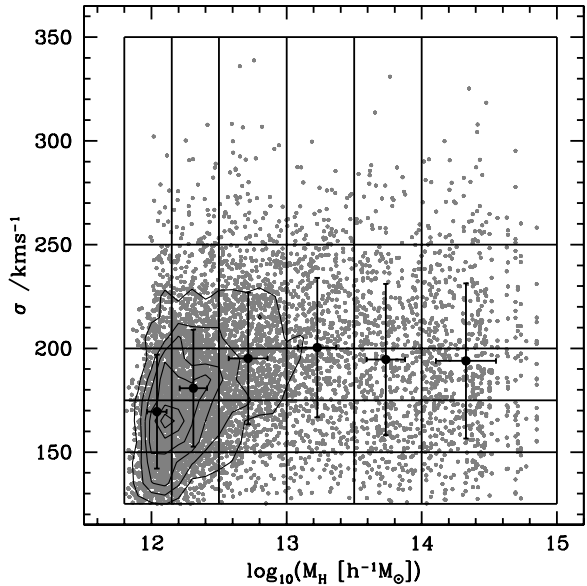}
  \end{center}
  \caption{Our sample of SDSS early-type galaxies is shown with the
    two main parameters that are used to characterise the intrinsic
    (velocity dispersion; $\sigma$) and environmental (host halo
    mass; M$_{\rm H}$) dependence of the underlying stellar
    populations.  In the crowded regions of the figure, contour lines
    track the density of galaxies in the plot.  The grid corresponds
    to the binning applied throughout the paper.  The large black dots and
    error bars give the average and RMS within each bin in halo mass.
    \label{fig:samp}
  }
\end{figure}

We make use of the Galaxy Groups Catalogue of \citet{yng07}, which is
an improved application of the \citet{yng05} halo-based galaxy group
finder to the New York University Value-Added Catalogue \citep{NYVAC},
based on the Sloan Digital Sky Survey Data Release 4
\citep{SDSStec}. We cross correlate this catalogue with the original
sample to find halo masses for all but 175 galaxies, leaving a total
of 6,959 galaxies in the new sample used here. We have also removed a
small ($\sim 150$) set of galaxies with velocity dispersions below
125~km/s, as these galaxies only appear in the bin with the lowest
halo masses, and have no counterparts in more massive halos. Allowing
them to be part of the lowest velocity dispersion bin would considerably
reduce the average $\sigma$ in this bin, introducing a bias when compared 
to other halo masses.

The galaxy group finder, described in \citet{yng05} and \citet{yng07},
is an iterative process in which the membership of the groups and the
relationships between the properties of the halo are refined at each
step. Initially the group finder uses a friends of friends algorithm
with small linking lengths to identify the centres of possible galaxy
group candidates. In such groups the centre of the halo is given by
the luminosity-weighted centre. The remaining isolated galaxies not
associated with groups are also set as the potential centres of
groups. The characteristic luminosity (L$_{19.5}$) of each candidate
group is then estimated, where this is defined as the summation of the
luminosity of all group members with $^{0.1}$M$_r - 5\log h\simlt -$19.5
\footnote{$^{0.1}M_r$ is the SDSS-r band magnitude for which K
and E corrections are applied at redshift z$=$0.1}. The L$_{19.5}$
values are corrected for survey completeness and the apparent
magnitude limit of the survey at redshifts z$\geq$0.09.

\begin{figure}
  \begin{center}
    \includegraphics[width=3.3in]{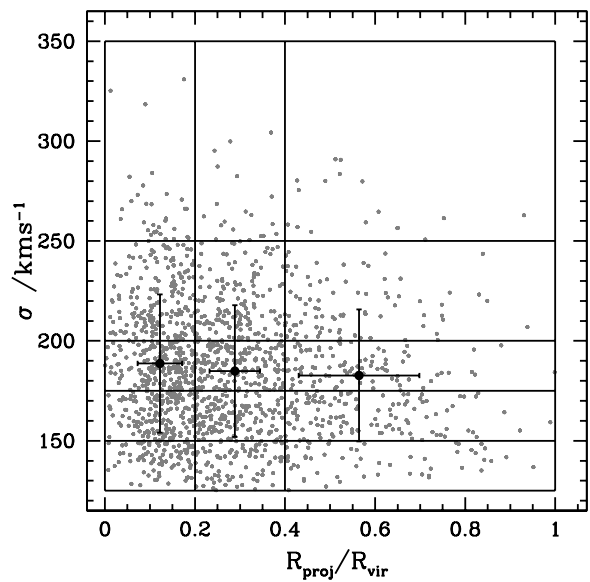}
  \end{center}
  \caption{The projected radial distance from the luminosity-weighted
    centre of the group is shown with respect to velocity
    dispersion. The distance is scaled to the virial radius, as determined from
    the properties of the halo.  The grid overlaying the sample shows
    the binning applied in the analysis. Black dots and error bars
    correspond to the average and RMS scatter of the sample within the
    bins in R$_{\rm proj}$.\label{fig:sampr} }
\end{figure}

Using the characteristic luminosity, the mass of the group halo is
estimated from a group mass-to-light ratio. The ratio used is set at a
constant value across all groups for the first iteration but is
subsequently refined to be group mass dependent. However, the results
are not particularly sensitive to the exact value of the ratio even if
it is held fixed \citep{yng05}. The estimate of the group halo mass
allows the derivation of other group halo properties such as the halo
radius, within which the halo has an average density contrast of 180,
and the virial radius, defined as the radius within which the average
density is above a set value.  Once the properties of the halo have
been estimated, a NFW profile is used for the dark matter
\citep{NFW97} to determine the three dimensional density contrast of
the halo in redshift space. Further galaxies are subsequently assigned
to the galaxy group candidates if they are within a certain distance
of the centre. This process -- which allows for the merging of two
groups if all members satisfy the above criteria singularly -- is
repeated until the membership of each group remains constant.  The
final dark matter halo mass is then estimated from a linear
relationship with respect to stellar mass, derived from semi-analytic
models \citep{kang05}. The galaxy group differentiates between central
and satellite galaxies. Those galaxies which are the most massive
galaxy of the group are defined as centrals, whereas the remaining
galaxies are labelled as satellites. Low mass halos can consist of a
single, central galaxy.
Shown in figure \ref{fig:samp} is the entire sample as a function of
host halo mass, M$_{\rm H}$, and central velocity dispersion,
$\sigma$, of the individual galaxies. The black dots correspond to the
average and RMS scatter of the sample, within the halo mass bins 
shown by the grid. 

\begin{figure*}
  \begin{minipage}{18cm}
    \includegraphics[width=3.5in]{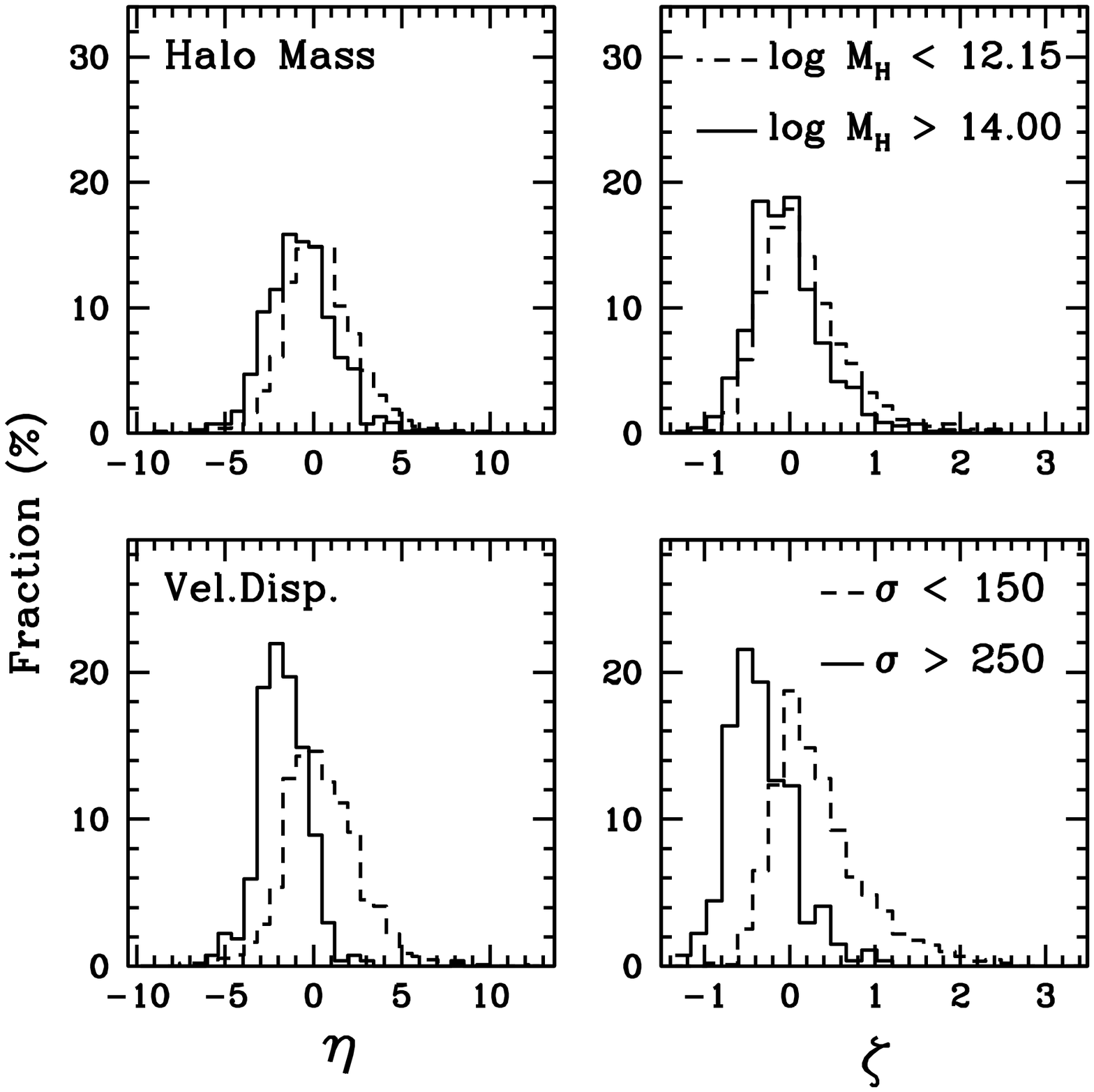}
    \includegraphics[width=3.5in]{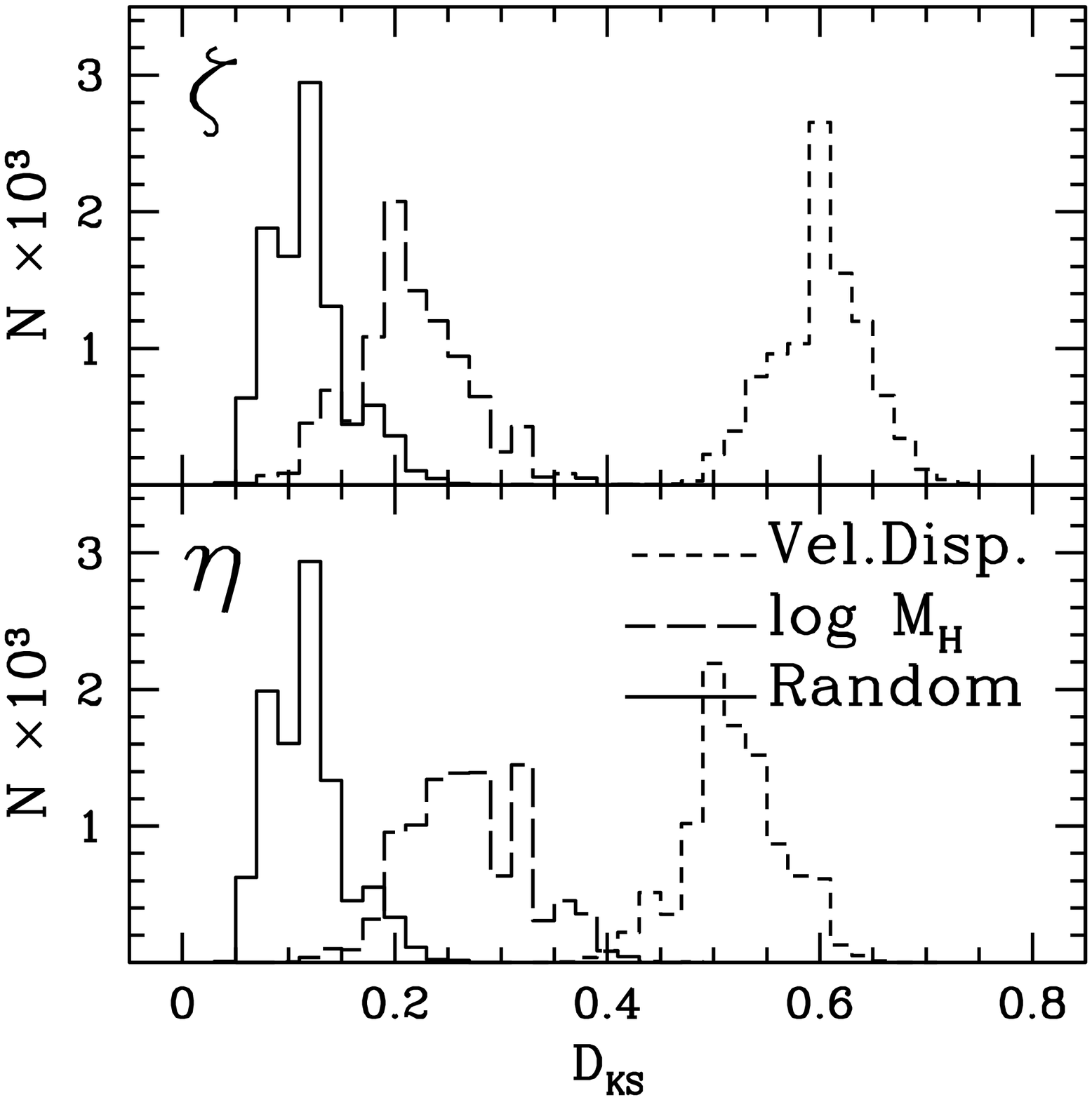}
    \caption{{\sl Left:} Distribution in PCA components $\eta$ and $\zeta$ of the
      upper and lower bins in velocity dispersion (\emph{bottom},
      labelled in km/s) and host halo mass (\emph{top}, labelled
      in $\log($M/M$_\odot)$) of the entire sample in $\eta$ and $\zeta$.  {\sl Right:} 
      distribution of the KS statistic (D$_{\rm KS}$) for the subsamples of
      the same bins in velocity dispersion (dashed) and halo mass (long-dashed) for
      $\eta$ (\emph{Bottom}) and $\zeta$ (\emph{Top}). In order to 
      quantify the statistical significance of the segregation, 
      the result of drawing subsamples at random over all mass ranges
      and environments is shown (solid).
      \label{fig:EZdist}}
  \end{minipage}
\end{figure*}

We can see from the figure that within the lowest halo mass bin 
(M$_{\rm H}\sim 10^{12}h^{-1}$M$_\odot$) there is a limitation imposed by 
the size of the halo on the highest sigma galaxy that it can contain. This 
is similar to that seen in the main catalogue in reference 
to the maximum stellar mass \citep[see e.g.][]{yang08}. The difference here 
is the non-trival mapping of velocity dispersion to dynamical or stellar
mass, which is dependent on other factors such as structure. However note that 
since the results presented in this paper are consistent 
across the range of velocity dispersions considered, the details of such 
mapping would be unlikely to affect them. Also we are using velocity dispersion as a 
measure of the intrinsic properties of the galaxy, which is consistent with many 
stellar population studies in the literature \citep[e.g.][]{bern03,gallazzi05}. 

The figure also reveals a shift towards decreasing velocity dispersion at lower 
halo mass, whereby less massive halos on average contain smaller galaxies.
This effect is illustrated by the solid circles and error bars in figure~\ref{fig:samp} 
which represent the average and root mean square deviation of the velocity 
dispersion within each bin of halo mass.

This correlation between mass / luminosity and environment in terms of
halo mass \citep{VDB08a} but also seen with respect to galaxy density
\citep{blanton05,croton05}, means that it is important to take into
consideration how such intrinsic properties change as a function of
environment. It is well known that the stellar populations of a galaxy
are related to its velocity dispersion
\citep{kauff04,thomas05,gallazzi05}, therefore such a correlation can
generate spurious environmental effects. Hence, in order to carry out
a robust analysis, it is important to separate out the effects of
environment, from those that are caused by this selection bias.

To achieve this, we investigate the difference with respect to environment
only at the same $\sigma$ range. Our sample is divided
into sub-samples either with respect to halo mass, or velocity dispersion.
This classification is shown as a grid in figure \ref{fig:samp}. The choice
of the grid size is motivated by the density of the underlying sample, 
selecting larger bins at high mass.

We also consider the variation in the properties of stellar
populations with respect to the position of a (satellite) galaxy
within its host halo. The dependence is measured as a function of the
projected distance from the luminosity weighted centre of the group,
namely the projected halo-centric radius, $R_{\rm proj}$. This will
allow us to assess the effect of galaxy accretion in groups on the
star formation history. We expect galaxies on the
outskirts of a group to be newer members. Hence, these galaxies will be
increasingly subject to the various effects of the group environment,
such as 'strangulation', 'ram-pressure stripping' and 'harassment'
which act on the satellite galaxies of the groups
\citep{wein06,VDB08a,VDB08b}.  Shown in figure
\ref{fig:sampr} is the satellite-only sample as a function of
projected distance scaled by R$_{\rm vir}$. 
Similar studies of satellites have found a radial mass segregation 
within groups \citep{VDB08a}, with the least massive galaxies at the 
outskirts of the group. We find no such trend within our sample with respect
to velocity dispersion.

\section{The stellar populations of different halo masses}

Regarding the effect of environment on the stellar populations of
elliptical galaxies, it is important to notice that the reported
differences have generally been small. For example, studies of the
colour-magnitude relation have only found limited and statistically
weak evidence \citep{bern03,gallazzi06}, even more comprehensive
analyses \citep{clemens06, bern06} find difference of the order $\sim
1$~Gyr.  In this paper we optimise the extraction of differences from
spectroscopic data via principal component analysis, which has been
shown to succeed in detecting differences within highly homogeneous
samples \citep[e.g.][]{igpca,pca}.

\subsection{PCA}
Principal Component Analysis (PCA) is a multivariate technique that
reduces the dimensionality of a data set. In most cases this is just
used as a data compression algorithm. However, previous work
\citep{mad03,igpca,pca} has shown that vital information can be
extracted from the projections on to the first few principal
components. In this work the variables that describe the data set are
the flux values at each wavelength i.e. the spectra.  The task of PCA
is to generate a set of basis vectors (the principal components) from
the data set, such that one can rank these vectors with respect to the
variance they capture. Hence, when obtaining the ``coordinates'' of
each galaxy by projecting their spectra on to the principal components,
one can use just the very few coordinates that correspond to the
principal components with the highest variance. In \citet{pca} we
find that the first two components already hold valuable information
regarding the average age of the stellar populations and the presence
of recent star formation. We refer the reader to that paper for
a detailed description of the methodology, although a brief summary 
is given below.

The first principal component is found to be consistent with a
typically old stellar population, showing a pronounced 4000\AA\ break,
significant metal absorption lines and limited Balmer line
strengths. The second component is much bluer, with absorption lines
dominated by the Balmer series. A correlation between both principal
components is forced by the orthogonality inherent to PCA. When
projecting onto the observed spectra, the components give a relation
that represents the mass-metallicity-age correlation of early-type
galaxies \citep[e.g.][]{bern03,thomas05}. The extended scatter in the
direction of the second component i.e. with respect to an excess of
blue light, is created by recent star formation. This issue is
confirmed in two ways: 1) through the application of a two component
stellar population model, where we find galaxies with a higher
projection onto the second principal component require a higher mass
fraction in young stars \citep{pca,cloprs}, 2) comparing the results of
PCA with NUV photometry from GALEX: The NUV$-$r colour is highly
sensitive to small amounts of recent star formation
\citep{Scha06,kav07}. Within our sample, 'NUV bright' galaxies
(NUV$-$r$\leq$4.9) present higher projections of the second principal
component, compared to the sample of quiescent, 'NUV faint' galaxies
(NUV$-$r$\geq$5.9). This trend between PCA and the presence of young
stars is optimised by rotating the projections on the two dimensional
plane spanned by PC1 and PC2, to give two new components: $\eta$ and
$\zeta$. These two components can be defined as the distance along the
PC1-PC2 correlation ($\eta$) and the residual of the correlation
($\zeta$). Even though NUV is more sensitive to the presence of
massive stars than optical spectra, this method is complementary to
NUV studies.  An advantage of using optical spectra over NUV
photometry is that we can track the presence of recent star formation
for a greater length of time: NUV light decays very rapidly as the
most massive stars die out.

In \citet{pca} we compare the PCA projections with a number of star
formation histories combined with population synthesis models and
conclude that $\eta$ is mainly related to the average age of the stellar
populations, whereas $\zeta$ tracks the presence of recent star formation
\citep[see also][]{PhD}.
This interpretation is carried over to this paper for the analysis
of the effects of environment.

\begin{figure}
  \begin{center}
    \includegraphics[width=3.4in]{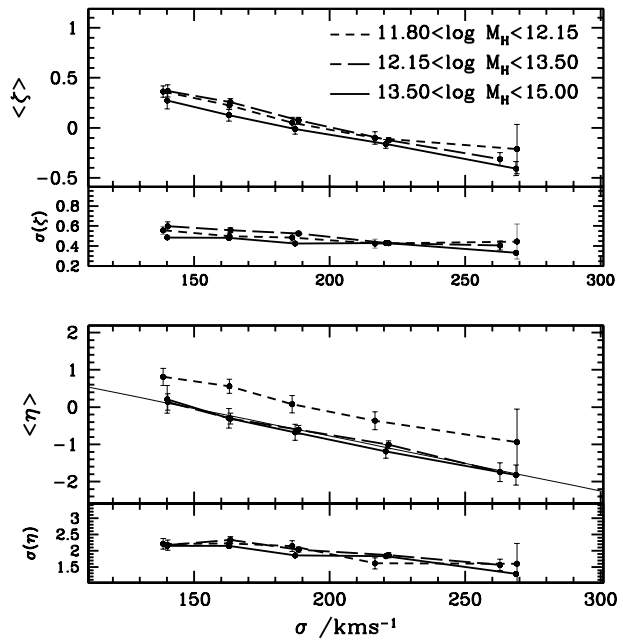}
  \end{center}
  \caption{The average of the $\eta$ and $\zeta$ components as a
    function of $\sigma$ within each group halo mass bin, as
    labelled (given in $\log($M/M$_\odot)$). The error bars correspond
    to the error on the mean. The small panels below the main plots
    show the standard deviation of $\eta$ and $\zeta$ in each bin. The
    thin black line is for reference only and is a least
    squares fit to the three most massive halos.\label{fig:HaloSigma1}
    }
\end{figure}

\subsection{Stellar Mass vs Group Halo Mass}
In this section we look into the dependence of the stellar properties
of elliptical galaxies on both the stellar mass and the mass of the
halo which the galaxy occupies. We utilise the results from PCA
described above, focusing on how the average properties of the galaxy
(through $\eta$) and the young stellar populations (through $\zeta$)
depend on galaxy mass or halo mass. While it is true that environment
plays a major role on the number density of early-type galaxies
\citep[i.e. the morphology-density relation, ][]{drez80}, here we pose
a different question: ``at a fixed $\sigma$, how different are the
star formation histories of early-type galaxies with respect to
environment?''. The left panel of figure~\ref{fig:EZdist} shows the distribution
of the first and last bins in both velocity dispersion ({\sl bottom}) and
group halo mass ({\sl top}). The comparison reveals that in terms of
the mass of the galaxy (i.e. velocity dispersion), smaller galaxies
have higher values of both $\eta$ and $\zeta$ indicating a younger
age, most likely due to increased recent star formation, in agreement
with the 'downsizing' scenario \citep{cowie96}. However the
distributions of the two extreme bins regarding environment 
({\sl top}) show a considerably smaller divergence in terms of their $\eta$
or $\zeta$ distributions.  Note that the preliminary results show that
those galaxies in the lowest mass halos have slightly higher values of
$\eta$ and $\zeta$.
 
\begin{figure}
  \includegraphics[width=3.5in]{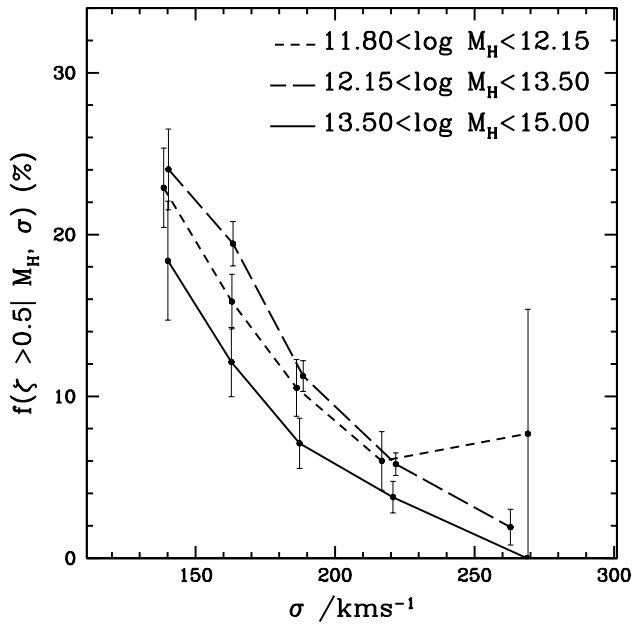}
  \caption{The fraction of galaxies with a $\zeta$ value above 0.5,
  which is consistent with the presence of recent star formation.
  Each line corresponds to a range of host halo masses, as labelled
  (given in $\log($M/M$_\odot)$). The
  error bars are given by Poisson errors.\label{fig:HMRSF} }
\end{figure}

The use of a Kolmogorov-Smirnov (KS) test confirms whether the
galaxies in these subsamples originate from the same
distribution. This is quantified in terms of the D$_{\rm KS}$
statistic \citep[a measure of the maximum difference of the cumulative
distributions, see e.g.][]{nr}. The right panel of
figure~\ref{fig:EZdist} shows the result of a Monte Carlo simulation,
where we extract samples of 100 galaxies from the upper and lower mass
bins, and perform the KS test in each case. This exercise is repeated
10,000 times, and the histogram of D$_{\rm KS}$ is shown when
extracting galaxies from the upper and lower bins in velocity
dispersion (short dashed line), in host halo mass (long dashed line) of
from random sampling of the complete sample (solid line).  Velocity
dispersion clearly plays the dominant role; its histograms in the
$\eta$ and $\zeta$ components are far from the distribution of random
sampling. However, environment -- or halo mass -- plays a more subtle
role, although a non-negligible one, especially with respect to the
$\eta$ component, which does show less of an overlap with the
distribution for random sampling.  Whether this effect is driven by
the mass bias mentioned above or whether is it a genuine effect of
environment is studied in this paper.

\subsection{A detailed look at the [$\eta$, $\zeta$] distributions}
Many authors have studied the role of environment on the scaling
relationships. This includes colours \citep[e.g.][]{wein06,kav07}, 
absorption line indices \citep[e.g][]{kun01,nelan05,bern06} as well 
as derived quantities such as luminosity-weighted age, metallicity 
and $[\alpha/$Fe$]$ \citep{thomas05,clemens06}. In a similar vein, 
we extend this analysis to include the $\eta$ and $\zeta$ components.

Shown in figure~\ref{fig:HaloSigma1} are the average values of $\eta$
and $\zeta$ with respect to velocity dispersion, for a range of host
halo masses.  Points corresponding to the same host halo mass range
are connected. The velocity dispersion value for each bin is given by
the average. A reference line is also plotted in the bottom panel.  It
is a least squares fit to the most massive halos and is shown in all
subsequent figures of $\langle\eta\rangle$ for comparison.
The error bars give the uncertainty on the mean values within each
bin. The RMS deviation is shown separately at the base of each panel
as $\sigma(\zeta)$ ({\sl top}) and $\sigma(\eta)$.

A clear dependence is found with respect to velocity dispersion, such
that galaxies with lower values of $\sigma$ have higher components
$\eta$ and $\zeta$, revealing the presence of younger stellar
populations.  Note that as shown in \citet{pca}, principal components
$\eta$ and $\zeta$ are not dependent on the velocity dispersion, and
so no correction is needed (in contrast with analyses of equivalent
widths).

It is important to notice from figure \ref{fig:HaloSigma1} that the
effect of environment is very subtle.  In terms of
$\langle\zeta\rangle$ and in most cases $\langle\eta\rangle$, the
trends of the different host halo mass ranges are indistinguishable
from each other. However, galaxies in the least massive halos show a
consistent and significantly higher value of $\langle\eta\rangle$,
across all values of $\sigma$ (although the nature of the trend is the
same for all halo mass bins). Given that the differences are small,
the significance is determined through a KS test for the $\eta$
component, performed by comparing galaxies in the lowest and the
highest host halo bin, across all values of $\sigma$.  In the three
central bins of velocity dispersion we find a probability higher than
99\% that the samples are drawn from different distributions.  At the
highest and lowest bins of $\sigma$ the small number of galaxies
prevent us from stating a similar result.

Furthermore, figure~\ref{fig:HaloSigma1} also indicates that
$\langle\zeta\rangle$ does not differentiate with respect to halo
mass.  This comes as a surprise, as one might have expected that
recent star formation is responsible for the higher values of
$\langle\eta\rangle$. To explore this issue in more detail, we use the
conditional fraction of galaxies with a value of $\zeta$ above a
threshold at which one needs to invoke recent star formation.  
We choose $\zeta\geq$0.5, since galaxies above this value both require 
significant fractions of young stars in a two-burst model and also 
have NUV luminosities consistent with recent star formation \citep{pca,cloprs}.

In figure~\ref{fig:HMRSF} we show this conditional fraction as a
function of velocity dispersion for a range of host halo masses.
Consistent with the simple analysis performed above, the main driving
force behind the fraction of galaxies with recent star formation is
galaxy mass (assuming $\sigma$ is representative of the mass of the 
galaxy). With respect to halo mass, the fraction of
younger galaxies seem to split such that there is a consistent lower
fraction for the most massive halo bins (black solid line)
relative to the rest of the sample. The significance of this
drop is not high in any one bin but the trend is notably consistent
across all bins. This is a qualitatively similar result to that
found in \cite{Scha06} using NUV data, although here we see that
recent star formation persists in halos below M$_{\rm H}\leq 3\times 10^{13}
h^{-1}$M$_\odot$ and is reduced in more massive halos. The possible
mechanisms underlying this effect are discussed in the conclusions.

\begin{figure}
  \begin{center}
    \includegraphics[width=3.5in]{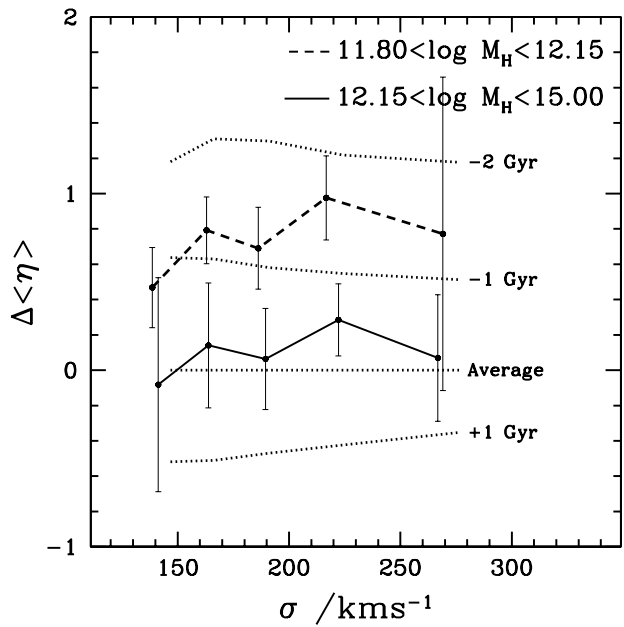}
  \end{center}
  \caption{Change in the $\eta$ component relative to the average
    relationship with $\sigma$ via SSP fitting (see text for
    details). A comparison of the lowest mass halo (dashed line) to
    the most massive halos (solid line) is shown relative to
    deviations from this relationship due to an increase/decrease in
    the average SSP age across $\sigma$ (dotted lines).\label{fig:HaloSigmaSSP}}
\end{figure}

\subsubsection{Modelling the $\eta$ component}

We now turn to quantifying the difference in $\langle\eta\rangle$ from
figure~\ref{fig:HaloSigma1} through the application of stellar
population synthesis models. Since the change of $\eta$ is not
accompanied by a change in $\zeta$ or the fractional change in
high-$\zeta$ galaxies, the shift is unlikely to be related to recent
star formation. Furthermore, this shift is relatively small, which
implies composite models, such as an exponentially decaying star
formation history, may well blur any discriminating effect. 
Therefore, while the modelling of early-types can be pushed beyond 
relatively simple formation histories \citep{bmc}, it is more robust 
in this case to consider simple stellar populations.

The observations are compared to synthetic spectra through the
equivalent widths of absorption line indices. The model spectra are
extracted from the simple stellar populations of \citet{BC03}, using a
\citet{chab03} initial mass function. A detailed grid of models is
constructed over a range of ages $\{1\cdots 14$~Gyr$\}$ and
metallicities $-1.0\leq \log(Z/Z_{\odot})\leq +0.35$, at solar
abundance ratios.  We constrain the population parameters with
multiple age sensitive absorption features: H$\beta$, H$\gamma$,
H$\delta$, as well as the 4300\AA\ absorption band (G4300) and the
metallicity indicator $[$MgFe$]$ \citep{GonzPhd}. The absorption
features are measured in each of the observed galaxy spectra, after
being smoothed to the highest velocity dispersion within the
respective bin.  The equivalent widths (EWs) are estimated using the
method outlined in \citet{bmc}, in which the pseudo-continuum is
determined by a boosted median value of the surrounding 
spectrum. The specific absorption lines are targeted using a 20\AA\
window. We use a standard maximum likelihood method to compare the
observed and model absorption features. The model EWs are obtained in
the same way, smoothing the spectra to the velocity dispersion
considered. For each galaxy we use the probability-weighted age and
metallicity, where the probability is defined as
P$(t,Z)\sim\exp^{-0.5\Delta\chi^2}$, which gives a robust estimate of
the SSP values \cite[e.g.][]{gallazzi05}. The best fits give a reduced
average of $\langle\chi_r^2\rangle = 1.3$.

The average age and metallicity from each of the halo mass and
velocity dispersion bins is used to investigate the difference seen in
the average $\eta$ value in an equivalent analysis to that shown in
figure~\ref{fig:HaloSigma1}.  However, we find no significant
differences between the predicted SSP ages or metallicities with
respect to host halo mass across all velocity dispersions. This
illustrates the power of PCA to identify small differences in the
spectra of elliptical galaxies, which might be difficult for
model-based methods to identify, especially at low S/N.  Therefore in
order to assess the (second order) effects of environment on $\eta$,
we consider small perturbations with respect to the (first order)
correlation with velocity dispersion.

We define a fiducial relation between velocity dispersion and
age/metallicity through the average parameters of each of the complete
velocity dispersion bins i.e. including all halo masses.  The SSP
parameters of this relation are then offset with respect to the
average age for each bin in velocity dispersion. The effect on the
$\eta$ component of this change (computed directly on the models) is
estimated and compared to the observed values -- defining for each
galaxy a $\Delta\eta$ as the difference between the $\eta$ value of
the galaxy and that of the model corresponding to the same velocity
dispersion. The model values of $\eta$ are derived in the same way as
for the observed values, i.e. via projection onto the principal
components followed by a rotation of the projected values. This rather
simple analysis enables us to quantify the change in $\eta$.  Notice
that we assume only a perturbation in age, as this is by far the most
dominant effect reported in the literature
\citep{bern03,thomas05,nelan05}.

While we are only interested in identifying the magnitude of the shift
of the lowest mass halo, as a check we compare our relationship of
metallicity and age with velocity dispersion to those found by other
authors. We quantify this relationship through the slope of a linear
fit to the fiducial relation, which was found to be: $\Delta\log
(Z/Z_{\odot})/\Delta\log (\sigma) =$ 0.68 and $\Delta\log ($Age
$/$Gyr$)/\Delta\log(\sigma) =$ 0.38 respectively. These values are
comparable to those reported in \citet{thomas05} (0.55, 0.24),
\citet{clemens06}(0.76,$\sim$) and \citet{nelan05}(0.53,0.59).

The $\sigma$ vs. $\Delta\eta$ relationship is shown in
figure~\ref{fig:HaloSigmaSSP} for the fiducial model obtained from the
best fit SSP ages and metallicities (labelled ``Average''). For reference,
we also show the relation when the age is shifted by 1 or 2 Gyr as labelled.
We include in the figure the observed values for the lowest halo mass
(dashed line) and for the rest of the sample (thick solid line).
The figure shows that early-type  galaxies in the lowest density regions
are on average about 1~Gyr younger than those in denser environments,
a result that is consistent for a wide range of velocity dispersion.
This result indicates that galaxies residing in all but the lowest
mass halos have similar stellar populations suggesting they are formed
in similar ways at similar redshifts. This may be due to a true
invariance across average/high density environments, such that above a
certain density the formation process becomes uniform.

\subsection{Centrals and Satellites}

The star formation histories of galaxies sitting at the centre of the
dark matter halos (centrals) and galaxies which orbit around the
centre (satellites) are expected to depend in a different way with
respect to the mass of their host halo.  Specifically,
satellites are more likely to be affected by the transformation
mechanisms operating in the halo environment \citep{VDB08a,VDB08b}. 
Therefore, we analyse separately the effects of environment on centrals 
and satellites and compare the relative differences.

\begin{figure}
  \begin{center}
    \includegraphics[width=3.5in]{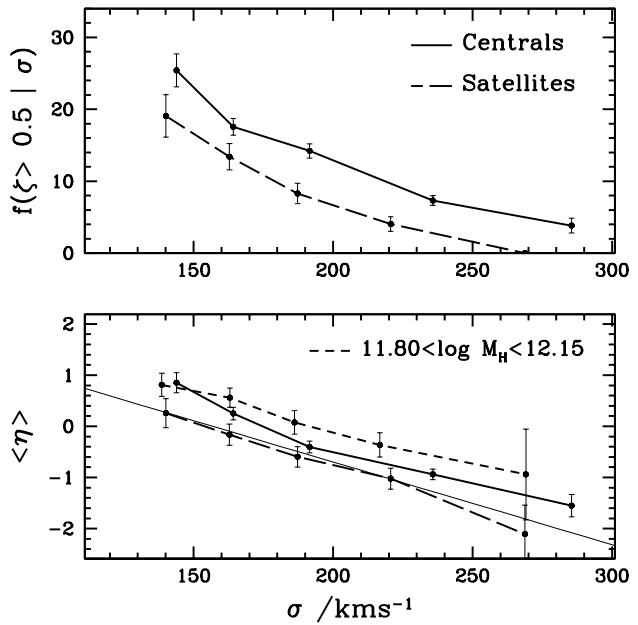}
  \end{center}
  \caption{Comparison of the distribution of the central and satellite
    galaxy populations at the same velocity dispersion, with respect
    to average age ($\langle\eta\rangle$; {\sl bottom}) and to recent
    star formation (f$(\zeta \geq 0.5)$; {\sl top}). The short dashed 
    line is the $\langle\eta\rangle$ for the lowest mass halo in the 
    whole sample. The thin solid line in the bottom panel is the reference 
    line from fig.\ref{fig:HaloSigma1}. \label{fig:satcen_V}
  }
\end{figure}

The comparison of satellite and central early-type galaxies at the
same velocity dispersion is shown in figure~\ref{fig:satcen_V}.  The figure
reveals that central galaxies (solid lines) have higher values of
$\langle\eta\rangle$ ({\sl bottom}) and f$(\zeta>0.5)$ ({\sl top}),
indicative of younger average ages \citep{pasq09b}.  and significant recent star
formation, respectively, when compared to satellites of the same mass
(long dashed lines).  This is consistent with \citet{VDB08b} and
\citet{wein09}, who found that in terms of optical colours,
centrals are bluer than satellite galaxies, indicating younger
ages. Here we can see this extends to the fraction of elliptical
galaxies which harbour small amounts of recent star formation as well.
However, notice that this trend includes all halo masses. In \S 3.4.1
we show the effect of separating this central/satellite classification
with respect to halo mass.

We have also plotted in the bottom panel of figure~\ref{fig:satcen_V}
the average value of $\eta$ for galaxies in the lowest mass halos
(short dashed line). At the lowest halo masses all galaxies are
central. Hence, this comparison allows us to rank the importance of
host halo mass against the central/satellite nature. The figure shows
that of these two properties, the mass of the halo dominates,
i.e. centrals in low mass halos are younger than centrals in
general. We focus on this aspect in the next two sub-sections.

\subsubsection{Effect on average age: $\eta$}

\begin{figure}
  \begin{center}
    \includegraphics[width=3.5in]{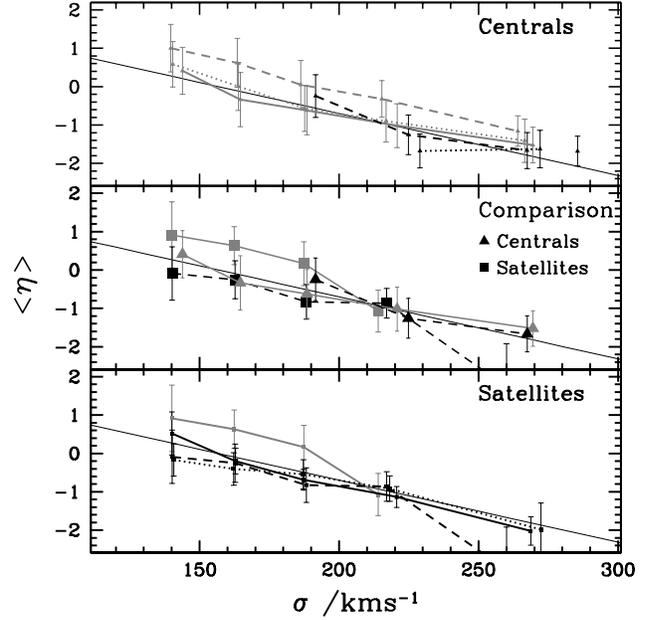}
  \end{center}
  \caption{$\langle\eta\rangle$ values for the central and satellite
    galaxy populations within each bin in velocity dispersion
    \emph{and} halo mass. The central panel compares central and
    satellite galaxies over a range of halo masses for which sufficient
    numbers of both types exist. The halo mass
    range is as follows -- given in $\log($M$_{\rm H}/$M$_\odot)$:
    grey dashed (11.80,12.15); grey dotted (12.15,12.50); grey
    solid (12.50,13.00); black dashed (13.00,13.50); black
    dotted (13.50,14.00); black solid (14.00,15.00). The thin solid line 
    shown in all panels is the reference line from fig.\ref{fig:HaloSigma1}.} 
  \label{fig:satcen_HMV}
\end{figure}

The average values of $\eta$ are plotted for the two subgroups in
figure~\ref{fig:satcen_HMV}. It is evident from the figure that the
analysis is complicated by the segregation imposed naturally by the
halo mass on the central/satellite nature of the galaxy. The sample
lacks central galaxies in the most massive halos (a limit imposed by
the survey volume). On the other hand, at a given velocity dispersion,
satellites cannot be found at low halo masses (a trivial constraint
imposed by the halo mass). This means a comparison of the two
populations can only occur within certain ranges of halo
mass. Therefore the middle panel of figure~\ref{fig:satcen_HMV} shows
the $\langle\eta\rangle$ values \emph{only} over a range of halo
masses in which a significant number of both centrals and
satellite galaxies exist. The top and bottom panels show the
central and satellite galaxies, respectively, over their full halo
mass range.

The central galaxies ({\sl top}), show an increase in the average
value of $\langle\eta\rangle$ at the lowest halo masses. This is
identical to the trend shown in figure~\ref{fig:HaloSigma1} (notice
centrals are the only type in the first halo mass bin). The result for
more massive halos is consistent with the main sample. This suggests
again that the effect of halo mass on the stellar populations is
limited to masses below M$_{\rm H}\sim 10^{12}$M$_\odot$. On the other
hand, satellite galaxies ({\sl bottom}) show an increase in
$\langle\eta\rangle$ in the group halo mass range $3\times 10^{12}
\leq$M$_{\rm H}\leq 10^{13}$M$_\odot$, which is the lowest halo mass
bin at which significant numbers of satellite galaxies exist. The
comparison in the middle panel reveals that at the overlapping halo
mass i.e. M$_{\rm H}\sim (3\times 10^{12} \cdots 10^{13})$M$_\odot$,
centrals have much lower values of $\langle\eta\rangle$. Thus the
increase of $\langle\eta\rangle$ found in the population of satellite
galaxies would have been hidden by the dominance of
centrals in this mass bin.

Given that higher $\eta$ values are consistent with younger average ages
suggests that satellites are in fact younger, when compared to central
galaxies of the same velocity dispersion and occupying halos of similar
mass \citep{pasq09b}. A KS test confirms that for this halo mass range, the second and
third bins in $\sigma$ of the centrals and satellites are drawn from
different distributions at a confidence level of $\geq$98\% and $\geq$99\%,
respectively. While this is seemingly in contrast with the previous
results, it is important to realise what is being compared in
figure~\ref{fig:satcen_HMV}. A pure central - satellite split (as in
figure~\ref{fig:satcen_V}) will generally result in a comparison of
central galaxies in low richness environments with satellites in 
higher density regions. This is the motivation of the split
in many papers, under the assumption that the centrals of today are a
reasonable approximation to the progenitors of the current satellite
population, (e.g. \S4.1 \citet{VDB08b}). Here we are comparing 
centrals in groups of significant size, to satellites in similar groups.
We find a trend in three out of the four bins considered,
although it is only found to be statistically significant in two. The
effect is not visible at the higher velocity dispersions.

\begin{figure}
  \begin{center}
    \includegraphics[width=3.5in]{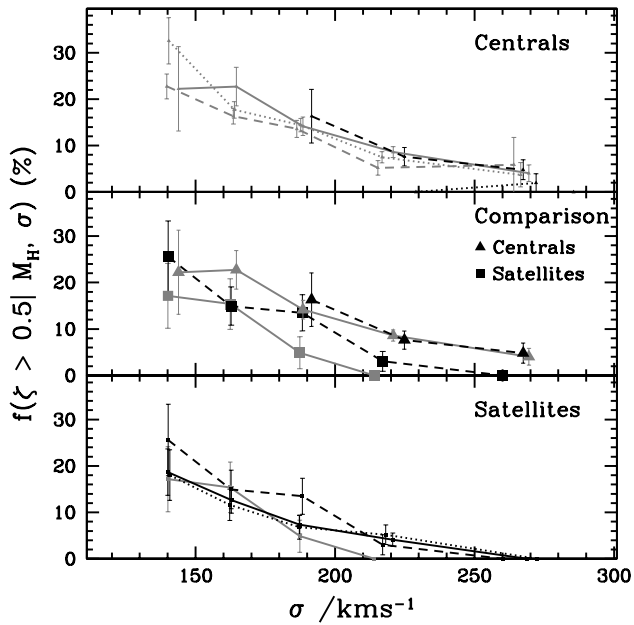}
  \end{center}
  \caption{Comparison of the distribution of central and satellite
    populations with respect to the fraction of galaxies with a
    $\zeta$ value above 0.5, which is consistent with recent star
    formation. The fraction is conditional i.e. calculated for each 
    bin separately.\label{fig:RSFsatcen}.  The binning with
    respect to halo mass is the same as in
    figure~\ref{fig:satcen_HMV}.  }
\end{figure}

Hence, we find that central galaxies 'evolve' faster than a similar
galaxy that is accreted onto a group. This probably relates to the
epochs over which environmental effects are efficient. Central
galaxies within more massive halos will have interacted with other
galaxies during the formation of the group and will also have existed
inside a more massive halo earlier than a galaxy of the same mass
which is accreted into a group halo. This result is consistent with the recent 
work of \citet{bern09}, who found that early-type brighest cluster 
galaxies were $\sim1$Gyr older than the surrounding satellite early-type 
galaxies.

Figure~\ref{fig:satcen_HMV} reveals that the satellite galaxies in
group halos of mass M$_{\rm H}\sim 3\times 10^{12} \cdots
10^{13}$M$_\odot$ have values of $\langle\eta\rangle$ consistent with
those of the central galaxies from halos of mass M$_{\rm H}\sim
6\times 10^{11} \cdots 10^{12}$M$_\odot$. This trend is consistent
with the hierarchical build up of structures expected within the
standard $\Lambda$CDM cosmology.

\subsubsection{Effect on recent star formation: $\zeta$}

Figure~\ref{fig:RSFsatcen} shows the fraction of galaxies with values
of $\zeta$ consistent with recent star formation, with respect to
central/satellite nature, in a similar manner to
figure~\ref{fig:satcen_HMV}. The trends of the individual populations
of centrals (\emph{top}) and satellites (\emph{bottom}) are consistent
with those found in the sample as a whole (see
figure~\ref{fig:HMRSF}). However a comparison of the fraction of high
$\zeta$ values between both centrals and satellites ({\sl middle})
shows a significant difference. The central galaxy data (triangles)
are consistently positioned at higher fractions of recent star
formation even within the same halo mass range. This might not be
particularly surprising: while satellites generally have their
accretion of material stopped when they enter the halo, centrals do
not, and so can still accrete gas. This mechanism is consistent with
the scenario of \citet{kav09} and \citet{cloprs}, whereby the recent
star formation seen in elliptical galaxies is fuelled by the accretion
of small clouds of gas or satellites,

\subsection{Effects with Halo-Centric Radius}

Since there is evidence to suggest that the younger low halo mass
galaxies are accreted as satellites, we might expect to see younger
galaxies on the outskirts of groups. There is also the possibility
that the halo-centric radius modulates the efficiency of environmental
mechanisms. We investigate this aspect by looking at
the relationship of $\eta$ and $\zeta$ as a function of the projected
distance from the luminosity weighted centre of the group.

\begin{figure*}
  \begin{minipage}{18cm}
    \includegraphics[width=3.4in]{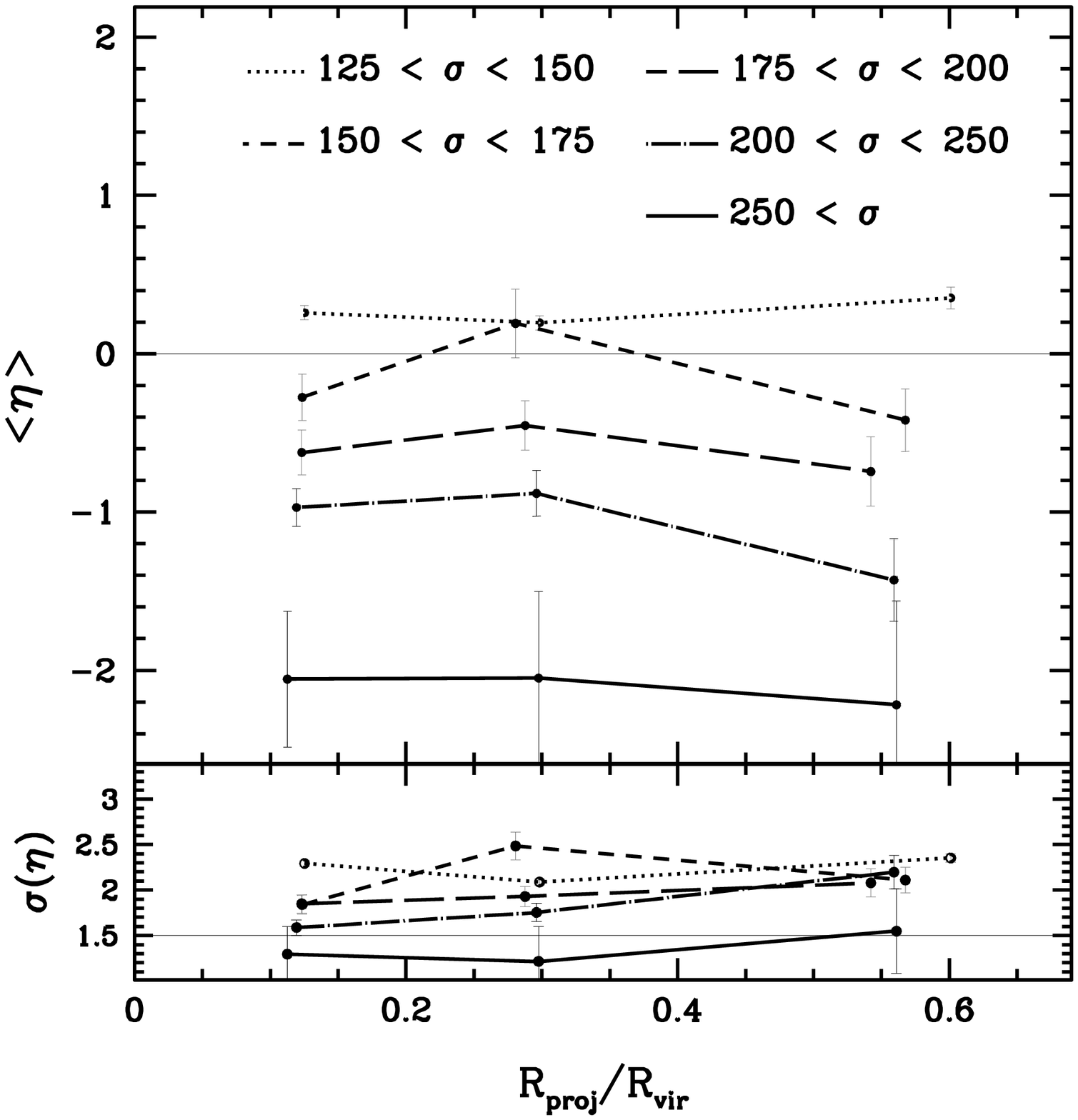}
    \includegraphics[width=3.4in]{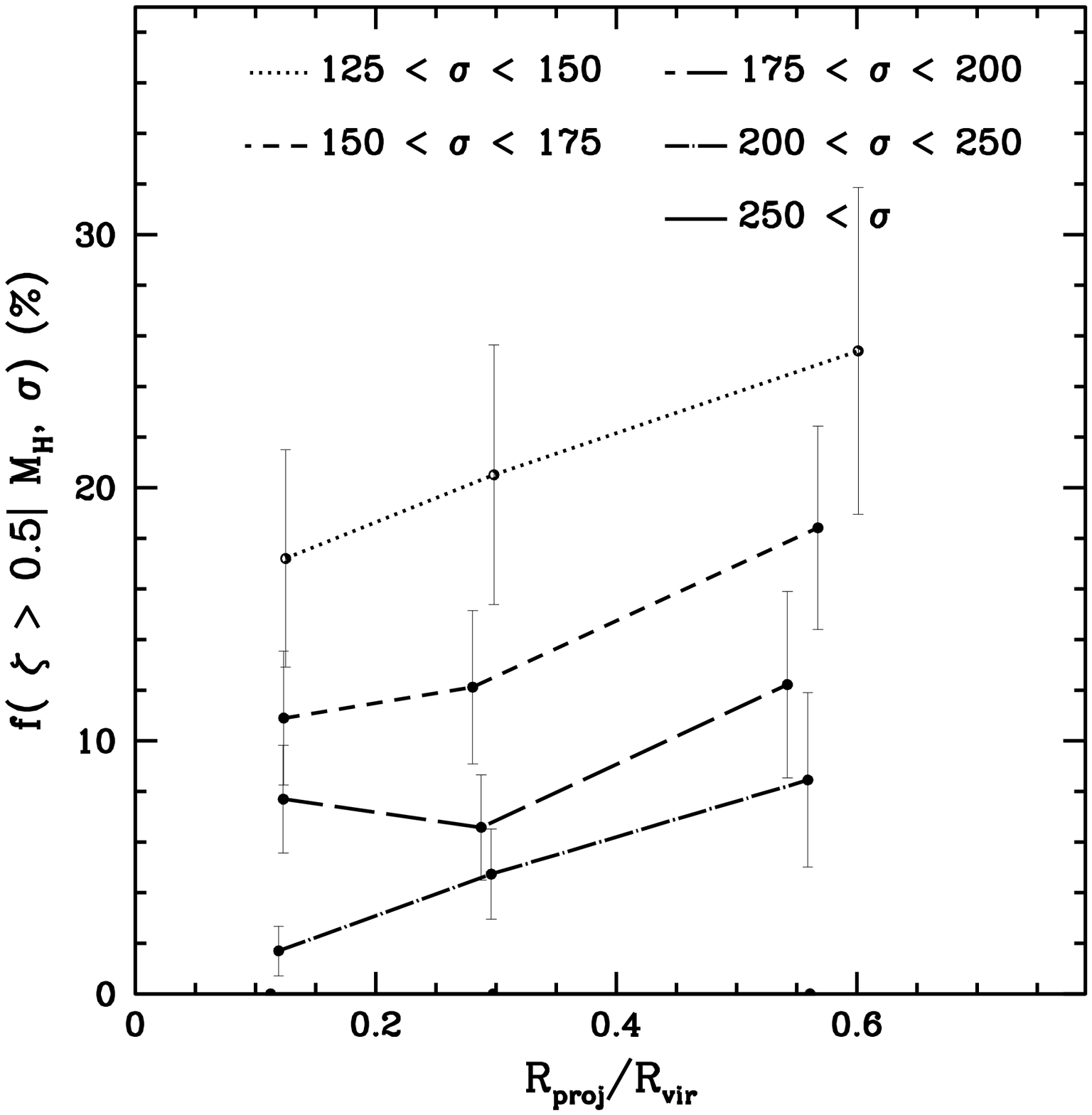}
    \caption{The $\eta$ and $\zeta$ components for the satellite
      subsample are shown as a function of distance from the
      luminosity weighted centre of the galaxy group, where the
      distance is scaled using R$_{\rm vir}$.  ({\sl Left:}) The main
      panel shows the average $\eta$ per bin in R$_{\rm proj}$/R$_{\rm
      vir}$, where adjoined points correspond to the same range in
      velocity dispersion (labelled in km/s). The smaller panel below
      shows the standard deviation within each bin. ({\sl Right:})
      Fraction of galaxies with a $\zeta$ value above 0.5 (i.e.  those
      with signatures of recent star formation). The fraction is computed
      within each bin separately. The error bars are the fractional Poisson
      errors.\label{fig:RV}}
  \end{minipage}
\end{figure*}

Shown in the right hand panel of figure~\ref{fig:RV} are the
satellite galaxies split into subsamples according to velocity
dispersion and halo centric radius, (see fig.~\ref{fig:sampr}),
against the average values of $\eta$ for each bin. Since we are
looking at galaxies across all group halo masses, the projected distance
is scaled using the virial radius of the occupied halo. As found by
previous work \cite[e.g.][]{VDB08a,wein09} there is no significant
trend. There is some indication that a reduction in the
scatter (bottom panel) of the $\eta$ values is seen for decreasing radii,
although the effect is obviously far from
robust. We note that this analysis is complicated both due to the
degenerate nature of the projected distance as well as to the fact
that smaller halos tend to be more concentrated. These effects are
likely to blur the subtle relationships associated with elliptical
galaxies and environment.

We also analyse in the left hand panel of figure~\ref{fig:RV} the
conditional fractions of galaxies which have $\zeta \geq 0.5$ as a
function of the scaled projected distance from the luminosity weighted
centre. An inspection of the figure indicates a general trend of
increasing recent star formation fractions with projected radius,
consistent with the previous comparison of centrals and
satellites. The stripping of a satellite surrounding gas may well
explain both the trend seen here and in
figure~\ref{fig:RSFsatcen}. Note that the recent star formation
fractions seen at the outskirts of the groups are higher than those
for the general satellite sample at the same stellar mass. This is
consistent with the results seen in \citet{igpca}; \citet{pca}; and
\citet{cloprs}, where we show that galaxies in medium density
environments show increased fractions of recent star formation
possibly due to interactions. The large error bars are a function of
the small numbers which exist away from the group centres, yet the
trend is consistent across all mass bins.

To conclude, we find that the effects with halo centric distance are
seen mainly in terms of $\zeta$ i.e. younger subpopulations.  As a
result of being accreted into the halo an elliptical galaxy will have
any residual star formation halted. On the other hand, we find that
the average age of the bulk of the stellar component does not depend
on the distance from the centre of the halo. This is to be expected
if, as argued by \cite{VDB08a}, a large fraction of early-type
galaxies transition onto the red sequence as centrals. It may be that
the main effects expected are better analysed within cluster samples
\cite[e.g.][]{nelan05}.

\section{Discussion and Conclusions}

Using the large $\sim$7,000 strong sample of early-type galaxies
described in \citet{pca} we have investigated the effect of
environment, measured through the mass of the dark matter halo that
hosts each galaxy. We make use of the previously derived PCA rotated
projections, $\eta$ and $\zeta$, sensitive to the average properties
and recent star formation, respectively, to identify small differences
in their stellar populations. We find that the star formation
histories are mainly a function of velocity dispersion. This is shown
in figure~\ref{fig:EZdist} where a considerable difference is evident
between the effects of velocity dispersion and group halo mass. When
we separate out the sample to remove the mass-environment degeneracy,
the stellar populations across most of the halos are
indistinguishable. Therefore, while the majority of this paper has
focused on the small differences observed in the elliptical galaxy
population, the main conclusion is that the effect of environment on
the stellar populations of early-type galaxies is 
{\sl extremely limited}. However, we do find that small but nonetheless interesting
differences can be detected with respect to environment.

First of all, galaxies within halos with the lowest masses
are estimated to be $\sim$1~Gyr younger than those in the rest of the
sample. This offset is consistent with the fact that galaxies in low mass
halos are exclusively centrals. Thus, they represent the lower end of the
environmental density scale and so our result compares favourably with
the general view that galaxies in low density regions are 1$-$2
Gyr younger \citep{bern98,thomas05,bern06,sanchezblaz06}.

We use in this paper a catalogue of groups to estimate the dark matter
halo masses, allowing us to compare more naturally with theoretical
modelling results. One of the more recent discoveries is that many
observations \citep[e.g.][]{binney04,croton05,dekel06,catt08} can be
well explained by assuming a critical halo mass above which the supply
of cold gas is stopped by virial shock heating
\citep{birn03,keres05,dekel06}. The critical halo mass is found to be M$_{\rm
H}\sim 10^{12}$M$_\odot$. Modelling by \citet{catt08} show that
galaxies above and below this value are consistent with the analysis
of \citet{thomas05} with respect to galaxies in high/low density
regions (see Cattaneo et~al. figure 6). It is possible that we see
this divide more explicitly here, since the lowest halo bin of this
sample is the only one to contain galaxies in halos below or close to
the critical mass halo considered here. Our sample gives a consistent
$\sim$1~Gyr age difference between galaxies in halos with masses above
and below this critical mass.  We do not find a gradual trend from the
lowest mass halo upwards, a result that may be caused by a combination
of effects. The limited timescale of evolutionary signatures on the
optical spectra \citep{harker06} and the importance of the individual
galaxy mass and the nature of the group build up or the low quality of
the data, may mean that the difference in SFH is only visible in the
lowest halo mass where the effect is strongest.

The average age is not the only parameter that varies across the
sample. The fraction of galaxies with a high value of $\zeta$
(consistent with recent star formation) is higher in the four lowest
mass halos. The reason for the drop above M$_{\rm H}\sim 3\times
10^{13}h^{-1}$M$_\odot$ is not entirely obvious. However, we note that
this is the same halo mass at which a decline is observed in the
optical AGN population \citep{gil07,pasq09}, possibly giving way to
radio-mode AGN. We also note that gas stripping processes are more
effective in more massive halos. The modelling results of
\citet{simha09} indicate that satellites of medium and low mass
i.e. those containing the majority of the recent star formation, have
their accretion stopped efficiently at halo masses M$_{\rm H} \sim
10^{14}$M$_\odot$.  We also observe a decrease of recent star
formation in the satellite population as a whole, as well as with
decreasing halo centric radius, suggesting that such process works
across all halo masses although with differing efficiency.

We also investigated the differing effects of environment on the
population of central and satellite galaxies.  These galaxies are
dominant in low and high mass halos, respectively, which implies a
limited overlap on the parameter space spanned by M$_{\rm H}$ and
$\sigma$. In the range of halo mass over which a comparison is
possible, the two populations were found to have similar stellar
populations at high velocity dispersion ($\sigma\geq$200 km/s).
However, at lower values ($150$ km/s $\leq\sigma\leq$ 200 km/s), 
$\langle\eta\rangle$ is higher for satellite galaxies, as
expected for younger ages. Hence, the stellar populations of central
galaxies were formed earlier.  Environmental effects preferentially
act on satellites whereas central galaxies mainly move onto the red
sequence as the result of a merger or when its halo mass surpasses the
critical mass of \citet{dekel06}.  Central galaxies generally sit at the peaks of
the dark matter density distribution \citep{berlind03}. In halos
with mass M$_{\rm H} \sim 3\times10^{13}h^{-1}$M$_\odot$, centrals are
within groups containing significant numbers of additional
(satellite) galaxies.  Therefore, they are likely to have been
subject to a higher level of interactions at earlier times and possibly 
had gas heated by infalling material and satellites \citep{KO_08}.
Furthermore, if we assume that the central galaxy is the core member of the group, we
would expect these galaxies to have crossed the threshold of
\citet{dekel06} earlier than a similar satellite, which would have
been accreted later. This would imply that at least {\sl some} satellites will
have become ellipticals after being accreted onto the group, creating
the lower mean ages. Therefore, we have the emerging picture that
while central galaxies are quenched on average at earlier times they
retain or accrete small amounts of gas with which to form small
amounts of stars.

\section*{Acknowledgments}
We would like to thank the referee, Sadegh Khochfar, for his useful
comments and suggestions. BR gratefully acknowledges funding from the
RAS. SK was supported by a Research Fellowship from the Royal
Commission for the Exhibition of 1851. We acknowledge use of the Delos
computer cluster at King's College London-Physics.  This work makes
use of the Sloan Digital Sky Survey. Funding for the SDSS and SDSS-II
has been provided by the Alfred P. Sloan Foundation, the Participating
Institutions, the National Science Foundation, the U.S.  Department of
Energy, the National Aeronautics and Space Administration, the
Japanese Monbukagakusho, the Max Planck Society, and the Higher
Education Funding Council for England. The SDSS Web Site is
http://www.sdss.org/.  The SDSS is managed by the Astrophysical
Research Consortium for the Participating Institutions.


\label{lastpage}
\end{document}